\documentclass[12pt,epsf]{article}
\usepackage{eurosym}
\usepackage{amsfonts}
\usepackage{amssymb,amsmath}
\usepackage{graphicx}
\usepackage[dvips]{color}
\usepackage{young}
\usepackage{youngtab}

\newcommand{\beq}{\begin{equation}}
\newcommand{\eeq}{\end{equation}}
\newcommand{\bea}{\begin{eqnarray}}
\newcommand{\eea}{\end{eqnarray}}
\newcommand{\beas}{\begin{eqnarray*}}
\newcommand{\eeas}{\end{eqnarray*}}
\newcommand{\ba}{\begin{array}}
\newcommand{\ea}{\end{array}}

\newcommand{\nbox}{{\,\lower0.9pt\vbox{\hrule \hbox{\vrule height 0.2 cm \hskip 0.19 cm \vrule height 0.2 cm}\hrule}\,}}

\def\href#1#2{#2}
\input{tcilatex}

\begin{document}

\begin{titlepage}
\hfill
\vbox{
    \halign{#\hfil         \cr
           } 
      }  

\hbox to \hsize{{}\hss \vtop{ \hbox{}

}}

%

\vspace*{20mm}
\begin{center}

{\large \textbf{$T^4$ fibrations over Calabi-Yau two-folds and non-K{\"a}hler \\
\vspace{0.2cm}
manifolds in string theory} }

{\Large \vspace{ 20mm} }

{\normalsize {Hai Lin}  }

{\normalsize \vspace{8mm} }

{\small \emph{\textit{Yau Mathematical Sciences Center, Tsinghua University, Beijing, 100084, P. R. China
}} }

{\normalsize \vspace{0.2cm} }

{\normalsize \vspace{0.4cm} }

%
\end{center}

\begin{abstract}
{\normalsize \vspace{0.2cm} }

We construct a geometric model of eight-dimensional manifolds and realize them in the context of type II string theory. These eight-manifolds are constructed by non-trivial $T^{4}$ fibrations over Calabi-Yau two-folds. These give rise to eight-dimensional non-K{\"{a}}hler Hermitian manifolds with $SU(4)$ structure. The eight-manifold is also a circle fibration over a seven-dimensional $G_{2}$ manifold with skew torsion. The eight-manifolds of this type appear as internal manifolds with $SU(4)$ structure in type IIB string theory with $F_{3}$ and $F_{7}$ fluxes. These manifolds have generalized calibrated cycles in the presence of fluxes.

\end{abstract}

\end{titlepage}

\section{Introduction}

\label{sec: introduction}

\vspace{1pt}

String theory has elegant and deep mathematical structures. It relates
theoretical physics to mathematics and has provided great insights to both
areas of research. In particular, a great number of important aspects of
geometric questions have occurred and can be addressed in the context of
string theory. For instance, manifolds with $SU(n)$ structure, such as the
Calabi-Yau $n$-folds, naturally appear in superstring theory and are
important subjects for our understanding.

\vspace{1pt}

An interesting model of manifolds with $SU(3)$ structure, is the geometric
construction of $T^{2}$ fibrations over Calabi-Yau two-folds \cite%
{Fu:2006vj,Goldstein:2002pg}. Such six-dimensional manifolds include not
only Calabi-Yau three-folds of the K\"{a}hler type, but also non-K\"{a}hler
Hermitian manifolds with $SU(3)$ structure. They can appear as the internal
six-manifolds when the superstring theory is compactified down to
four-dimensional spacetime. A natural question that is addressed by this
present paper is what happens if we use $T^{4}$ fibrations, instead of $%
T^{2} $ fibrations. This corresponds to a geometric model of
eight-dimensional manifolds that we construct in this paper.

\vspace{1pt}

Internal manifolds with six dimensions have been well-studied, in the
context of string compactification. However, eight-dimensional internal
manifolds are also very interesting. They can have similar mathematical
structures as their six-dimensional counterparts, for example they can be
Hermitian and have an $SU(n)$ structure where $n$ is the complex dimension.
Furthermore, balanced Hermitian manifolds exist in both six dimensions and
eight dimensions. Moreover, eight-manifolds can naturally appear in the
compactification of string theory with fluxes to two-dimensional spacetime.

\vspace{1pt}

Eight-dimensional manifolds with $SU(4)$ structure include both K\"{a}hler
Calabi-Yau four-folds and non-K\"{a}hler Hermitian manifolds with $SU(4)$
structure. These manifolds are equipped with a Hermitian two-form and a
holomorphic four-form. These forms can be constructed by bilinears of
internal Killing spinors. These eight-dimensional manifolds have been
studied by using the equations of pure spinors in type II string theory \cite%
{Prins:2013koa,Rosa:2013lwa,Prins:2013wza}. The K\"{a}hler Calabi-Yau
four-folds are the special cases, when both the Hermitian form and
holomorphic form are closed. In the presence of fluxes, these forms need not
be closed, and this is the case for the non-K\"{a}hler $SU(4)$-structure
manifolds.

\vspace{1pt}

The non-K\"{a}hler manifolds can appear naturally in string theory with
fluxes. In the compactification of heterotic string theory to four
dimensional Minkowski spacetime \cite{Candelas:1985en}, the internal
six-manifolds can become non-K\"{a}hler in the presence of fluxes \cite%
{Strominger:1986uh,Becker:2006et,Fu:2006vj,Becker:2009df}. Various models of
constructing heterotic manifolds and their vector-bundles have been put
forward [7-13]. They play an important role in searching for realistic
string theory vacua with four dimensional Minkowski spacetime.

\vspace{1pt}

An interesting type of non-K\"{a}hler manifolds, which are very important in
differential geometry, are balanced Hermitian manifolds. They are Hermitian
manifolds with a Hermitian form and a holomorphic form. For a balanced
manifold, unlike K\"{a}hler manifolds, its Hermitian form is not closed,
however, the ($n-1$)th power of its Hermitian form is closed, where $n$ is
the complex dimension of the manifold \cite{Mic82}. Since they impose a
weaker condition on the closure of the Hermitian form than the K\"{a}hler
manifolds, they represent close variants of K\"{a}hler manifolds. Some non-K%
\"{a}hler Hermitian balanced manifolds can have trivial canonical bundle,
and thus are interesting examples of non-K\"{a}hler Calabi-Yau manifolds,
see for instance \cite{Tosatti}. Moreover, under appropriate blowing-downs
or contractions of curves, some classes of balanced manifolds can become K%
\"{a}hler and have projective models in algebraic geometry.

\vspace{1pt}

In this paper, we will construct eight-dimensional manifolds of the non-K%
\"{a}hler Hermitian type, by $T^{4}$ fibrations over Calabi-Yau two-folds.
They have $SU(4)$ structures but are not the standard K\"{a}hler Calabi-Yau
four-folds. The eight-manifolds can also be viewed as a circle bundle over a
seven-dimensional base. We will show that the base is a $G_{2}$ manifold
with skew torsion. General $G_{2}$ manifolds with torsion have been widely
studied [16-20]. The geometric model of the eight-manifolds here, fits with
type II string theory with $F_{3}$ and $F_{7}$ fluxes and dilaton, as we
will see in the later sections.

\vspace{1pt}

The organization of this paper is as follows. In Sec. \ref{sec: T4}, we
construct eight dimensional Hermitian manifolds\ by $T^{4}$ fibrations over
Calabi-Yau two-folds. In Sec. \ref{sec: G2}, we find that the eight-manifold
of this type can be viewed as a circle bundle over a seven-dimensional $%
G_{2} $ manifold with skew torsion. After that in Sec. \ref{sec: type II
string theory}, we find that the eight-manifold of this kind can be used in
type IIB string theory on the warped product of a two-dimensional Minkowski
spacetime and an eight-manifold. Then in Sec. \ref{sec: generalized
calibrated cycles}, generalized calibration forms and generalized calibrated
cycles are constructed for these models appearing in the type IIB string
theory. Finally we briefly discuss related aspects in Sec. \ref{sec:
discussion}.

\vspace{1pt}

\vspace{1pt}

\vspace{1pt}

\section{$T^{4}$ fibrations over Calabi-Yau two-folds and non-K{\"a}hler
eight-manifolds}

\label{sec: T4}

In this section we construct a geometric model of eight-dimensional
Hermitian manifolds, by fibrations of four-dimensional tori $T^{4}$ over
four-dimensional base manifolds which are complex. We devote particular
attention to the case that the four dimensional base is a Calabi-Yau
two-fold.

Let us consider a ten-dimensional metric of string theory arising as a
warped product of a two-dimensional Minkowski spacetime $R^{1,1}$ and an
eight-dimensional manifold $M^{8}$. The line element of the ten-dimensional
metric is
\begin{equation}
ds^{2}=e^{2A}ds^{2}(R^{1,1})+ds^{2}(M^{8}).  \label{M10_metric_01}
\end{equation}%
Here, $M^{8}$ is a non-trivial $T^{4}$ fibration over a four-manifold $M^{4}$%
\begin{equation}
T^{4}\overset{}{\rightarrow }M^{8}\overset{}{\rightarrow }M^{4}.
\end{equation}%
We define the projection map
\begin{equation}
\pi :M^{8}\overset{}{\rightarrow }M^{4}.
\end{equation}%
In general, we can consider the eight-dimensional manifold $M^{8}$ to be
either compact or non-compact. For instance, we can obtain non-compact $%
M^{8} $ by taking the base $M^{4}$ to be non-compact. The $e^{2A}$ in the
metric (\ref{M10_metric_01}) is a warp factor in front of the metric of $%
R^{1,1}$.

The line element of the eight-dimensional metric is
\begin{equation}
ds^{2}(M^{8})=e^{2v}[\func{Re}(\theta _{(1)}\otimes {\bar{\theta}}%
_{(1)}+\theta _{(2)}\otimes {\bar{\theta}}_{(2)})+e^{2C}ds^{2}(M^{4})],
\label{M8_metric_01}
\end{equation}%
where%
\begin{eqnarray}
\theta _{(1)} &=&dx_{1}+idy_{1}+A_{(1)}, \\
\theta _{(2)} &=&dx_{2}+idy_{2}+A_{(2)}.
\end{eqnarray}%
We consider $M^{4}$ as a complex manifold, equipped with a Hermitian
two-form $J_{M^{4}}$ and a holomorphic two-form $\Omega _{M^{4}}$, so that $%
d\Omega _{M^{4}}=0$. The $\{x_{1},y_{1},x_{2},y_{2}\}$ are coordinates of
the tori $T^{4}.$ The connections of the fibrations are complex one-forms $%
A_{(1)}$ and $A_{(2)}.~$Their curvatures are $F_{(i)}=dA_{(i)}$ and $\bar{F}%
_{(i)}=d{\bar{A}}_{(i)},$ for $i=1,2$.$~$The $e^{2A},e^{2v},~e^{2v+2C}$ are
three warped factors. The $A,v,C$ are functions on the four-manifold $%
M^{4}.~ $The function $e^{2C}$ is a warp factor in front of the metric of $%
M^{4}.$

The line element of the $M^{4}$ can be written as
\begin{equation}
ds^{2}({\tilde{M}}^{4})=e^{2C}ds^{2}(M^{4}),
\end{equation}%
where ${\tilde{M}}^{4}$ is a Hermitian manifold with $J_{{\tilde{M}}%
^{4}}=e^{2C}J_{M^{4}}$~and~$\Omega _{{\tilde{M}}^{4}}=e^{2C}\Omega _{M^{4}}$.

Now let us describe the geometry of the $T^{4}$ fibration in more detail.
The $A_{(1)}$ and $A_{(2)}$ are the pull-backs of the complex one-forms $%
a_{(1)}$ and $a_{(2)}$ on the base complex four-manifold $M^{4}$. In other
words, $A_{(1)}=\pi ^{\ast }a_{(1)}$ and $A_{(2)}=\pi ^{\ast }a_{(2)}$. We
assume that $a_{(1)}$ and $a_{(2)}$ are of the $(1,0)$ type on the base. In
component form, the fibration of the $T^{4}$ is described by
\begin{eqnarray}
&&(dx_{1}+\pi ^{\ast }\func{Re}a_{(1)})^{2}+(dy_{1}+\pi ^{\ast }\func{Im}%
a_{(1)})^{2}+(dx_{2}+\pi ^{\ast }\func{Re}a_{(2)})^{2}+(dy_{2}+\pi ^{\ast }%
\func{Im}a_{(2)})^{2}.  \notag \\
&&
\end{eqnarray}%
The curvatures $f_{(1)}$~and $f_{(2)}$ of the complex one-forms on the base
can be written locally as $f_{(1)}=da_{(1)}$ and $f_{(2)}=da_{(2)}$. The
connections $a_{(i)}$ and ${\bar{a}}_{(i)}$, for $i=1,2$, have curvatures
such that $[-\frac{f_{(i)}}{2\pi }],[-\frac{{\bar{f}}_{(i)}}{2\pi }]\in
H^{2}(M^{4},\mathbb{Z})$. We see that $F_{(i)}=\pi ^{\ast }f_{(i)}$ and$~%
\bar{F}_{(i)}=\pi ^{\ast }{\bar{f}}_{(i)}.~$

The eight-manifold $M^{8}$ is hence a Hermitian manifold equipped with a
Riemannian metric in (\ref{M8_metric_01}), a Hermitian $(1,1)$ form $J$, and
a holomorphic $(4,0)$ form $\Omega $. Let us denote
\begin{equation}
J_{(1)}=\frac{i}{2}\theta _{(1)}\wedge {\bar{\theta}}_{(1)}\,,~~~~J_{(2)}=%
\frac{i}{2}\theta _{(2)}\wedge {\bar{\theta}}_{(2)}\,.
\end{equation}%
The Hermitian form $J$ and the holomorphic form $\Omega $ of $M^{8}$ are%
\begin{equation}
J=~e^{2v}J_{(1)}+e^{2v}J_{(2)}+e^{2(v+C)}\pi ^{\ast }J_{M^{4}},  \label{J_01}
\end{equation}%
\begin{equation}
\Omega =e^{4v+2C}\theta _{(1)}\wedge \theta _{(2)}\wedge \pi ^{\ast }\Omega
_{M^{4}}.  \label{Omega_01}
\end{equation}%
We have that $J_{(1)}^{2}=0,J_{(2)}^{2}=0,\pi ^{\ast }J_{M^{4}}^{3}=0.~$The
holomorphic $(4,0)$ form $\Omega $ requires that $\theta _{(1)}$ and $\theta
_{(2)}$, and hence $A_{(1)}$ and $A_{(2)}$, are of the $(1,0)$ type. This is
the reason that we have assumed that $a_{(1)}$ and $a_{(2)}$ are of the $%
(1,0)$ type on the base.

Let us consider the closure properties of $\Omega $ and $J$. We first
analyze $\Omega,$
\begin{equation}
d(e^{-4v-2C}\Omega )=(F_{(1)}\wedge \theta _{(2)}-F_{(2)}\wedge \theta
_{(1)})\wedge \pi ^{\ast }\Omega _{M^{4}}.~  \label{Omega_02}
\end{equation}%
By demanding the vanishing of the right-hand side of Eq. (\ref{Omega_02}),
we assume the condition
\begin{equation}
F_{(i)}\wedge \pi ^{\ast }\Omega _{M^{4}}=0.
\end{equation}%
Hence, with this condition
\begin{equation}
d(e^{-4v-2C}\Omega )=0.~  \label{holomorphic_01}
\end{equation}

Now let us consider the closure property of $J,$%
\begin{eqnarray}
dJ &=&e^{2v}(\frac{i}{2}F_{(1)}\wedge {\bar{\theta}}_{(1)}-\frac{i}{2}\bar{F}%
_{(1)}\wedge \theta _{(1)}+\frac{i}{2}F_{(2)}\wedge {\bar{\theta}}_{(2)}-%
\frac{i}{2}\bar{F}_{(2)}\wedge \theta _{(2)})  \notag \\
&&+2dv\wedge J+2e^{2v+2C}dC\wedge \pi ^{\ast }J_{M^{4}}+e^{2v+2C}\pi ^{\ast
}dJ_{M^{4}}.  \label{dJ}
\end{eqnarray}%
Due to the presence of nonzero $F_{(1)},\bar{F}_{(1)},F_{(2)},\bar{F}_{(2)},$
the first line can not vanish. In other words,
\begin{eqnarray}
&&\frac{i}{2}F_{(1)}\wedge {\bar{\theta}}_{(1)}-\frac{i}{2}\bar{F}%
_{(1)}\wedge \theta _{(1)}+\frac{i}{2}F_{(2)}\wedge {\bar{\theta}}_{(2)}-%
\frac{i}{2}\bar{F}_{(2)}\wedge \theta _{(2)}  \notag \\
&=&-\func{Im}(F_{(1)}\wedge {\bar{\theta}}_{(1)}+F_{(2)}\wedge {\bar{\theta}}%
_{(2)})\neq 0.
\end{eqnarray}%
Therefore $J$ is not closed or conformally closed. Hence, with the nonzero $%
F_{(i)},\bar{F}_{(i)}$,$~$the eight-manifold $M^{8}~$is not K\"{a}hler and
not conformally K\"{a}hler.

If $M^{4}$ is complex and non-K\"{a}hler, then the non-K\"{a}hlerity of $%
M^{8}~$can be attributed to the base $M^{4}$ being non-K\"{a}hler, as from
the last term in (\ref{dJ}). To analyze situations when the non-K\"{a}%
hlerity of $M^{8}~$is not attributed to the base $M^{4}$ being non-K\"{a}%
hler, we consider $M^{4}$ being K\"{a}hler, in other words,
\begin{equation}
dJ_{M^{4}}=0.  \label{dJ_M4_01}
\end{equation}%
${\tilde{M}}^{4}$ is hence conformally K\"{a}hler.

Let us now consider
\begin{eqnarray}
d(J^{2}) &=&-\frac{1}{2}e^{4v}(F_{(1)}\wedge \theta _{(2)}\wedge {\bar{\theta%
}}_{(2)}\wedge {\bar{\theta}}_{(1)}-\bar{F}_{(1)}\wedge \theta _{(2)}\wedge {%
\bar{\theta}}_{(2)}\wedge \theta _{(1)}  \notag \\
&&+F_{(2)}\wedge \theta _{(1)}\wedge {\bar{\theta}}_{(1)}\wedge {\bar{\theta}%
}_{(2)}-\bar{F}_{(2)}\wedge \theta _{(1)}\wedge {\bar{\theta}}_{(1)}\wedge
\theta _{(2)})  \notag \\
&&+4dv\wedge J^{2}+4e^{4v+2C}dC\wedge \pi ^{\ast }J_{M^{4}}\wedge
(J_{(1)}+J_{(2)}),  \label{dJ^2}
\end{eqnarray}%
where we have used $dJ_{M^{4}}=0$ and assumed the condition%
\begin{equation}
F_{(i)}\wedge \pi ^{\ast }J_{M^{4}}=0,~~~~\bar{F}_{(i)}\wedge \pi ^{\ast
}J_{M^{4}}=0.  \label{F_i_02}
\end{equation}%
Due to the nonzero $F_{(1)},\bar{F}_{(1)},F_{(2)},\bar{F}_{(2)}$,$~$the $%
J^{2}$ is not closed or conformally closed.

Finally let us consider $dJ^{3}$. With the conditions (\ref{dJ_M4_01}) and (%
\ref{F_i_02}),
\begin{equation}
d(e^{-6v-2C}J^{3})=0.  \label{conformally_balanced}
\end{equation}%
Hence in this case $J^{3}$ is a conformally closed $(3,3)$ form.

The eight-manifold $M^{8}~$can be written as
\begin{eqnarray}
ds^{2}(M^{8}) &=&e^{2v}ds^{2}({\tilde{M}}^{8}), \\
ds^{2}({\tilde{M}}^{8}) &=&\func{Re}(\theta _{(1)}\otimes {\bar{\theta}}%
_{(1)}+\theta _{(2)}\otimes {\bar{\theta}}_{(2)})+e^{2C}ds^{2}(M^{4}),
\end{eqnarray}%
where $M^{8}$ is conformal to ${\tilde{M}}^{8}.$ The ${\tilde{M}}^{8}$ has a
Hermitian two-form ${\tilde{J}}$ and a holomorphic four-form ${\tilde{\Omega}%
~}$as follows,
\begin{equation}
{\tilde{J}}=~J_{(1)}+J_{(2)}+e^{2C}\pi ^{\ast }J_{M^{4}},  \label{J_t}
\end{equation}%
\begin{equation}
{\tilde{\Omega}}=\theta _{(1)}\wedge \theta _{(2)}\wedge \pi ^{\ast }\Omega
_{M^{4}}.  \label{Omega_t}
\end{equation}%
The norm of ${\tilde{\Omega}}$ with respect to the Hermitian form ${\tilde{J}%
}$ is%
\begin{equation}
\parallel {\tilde{\Omega}}\parallel _{{\tilde{J}}}=e^{-2C}.
\label{holomorphic_norm_}
\end{equation}%
We have that $d(e^{-2C}{\tilde{J}}^{3})=0,$ and from Eq. (\ref%
{holomorphic_norm_}), we see that
\begin{equation}
d(\parallel {\tilde{\Omega}}\parallel _{{\tilde{J}}}{\tilde{J}}^{3})=0.
\label{hermitian_02}
\end{equation}%
This expression (\ref{hermitian_02}) is for the ansatz in Eqs. (\ref{J_t})
and (\ref{Omega_t}). We have assumed that $M^{4}$ is K\"{a}hler in the above
derivation of Eq. (\ref{hermitian_02}). In order that the holomorphic
four-form ${\tilde{\Omega}}$ is non-vanishing, according to Eq. (\ref%
{Omega_t}), $M^{4}$ has a non-vanishing holomorphic two-form. Hence, by the
classification of complex surfaces by Enriques and Kodaira, $M^{4}$ are
Calabi-Yau two-folds. Under a conformal transformation, let ${\tilde{J}}%
^{\prime }=e^{-\frac{2C}{3}}{\tilde{J}}$,$~$then $d{\tilde{J}}^{\prime 3}=0$%
. This is the condition for eight-dimensional conformally balanced manifolds
\cite{Mic82}. Hence, $M^{8}$ is conformally balanced, with the additional
assumption used in the above derivation%
\begin{eqnarray}
F_{(i)}\wedge \pi ^{\ast }\Omega _{M^{4}} &=&0,  \notag \\
F_{(i)}\wedge \pi ^{\ast }J_{M^{4}} &=&0,  \label{condition_} \\
\bar{F}_{(i)}\wedge \pi ^{\ast }J_{M^{4}} &=&0.  \notag
\end{eqnarray}

The balanced manifolds have certain nice properties. Some balanced
manifolds, although not K\"{a}hler, after performing appropriate
blowing-downs or contractions of curves, have a limit that become projective
and K\"{a}hler, see for example \cite{Poon86,Lebrun
Poon,Poon,Lin:2014lya,Fu:2008zh}. Some smooth balanced manifolds can appear
as crepant resolutions of certain projective and K\"{a}hler manifolds, see
for example the six dimensional case discussed in \cite{Poon86,Lebrun
Poon,Poon,Lin:2014lya}.

Now let us consider what the condition (\ref{condition_}) imply for the base
$M^{4}~$and the fibrations of $T^{4}$. \vspace{1pt}The space of the
two-forms on $M^{4}$ can be decomposed by the direct sum of the space of
self-dual two-forms $\Omega _{M^{4}}^{2+}~$and the space of anti-self-dual
two-forms $\Omega _{M^{4}}^{2-}$,
\begin{equation}
\Omega _{M^{4}}^{2}=\Omega _{M^{4}}^{2+}\oplus \Omega _{M^{4}}^{2-}.
\end{equation}%
For a complex manifold $M^{4}$, it can be further decomposed as%
\begin{equation}
\Omega _{M^{4}}^{2+}=\Omega _{M^{4}}^{2,0}\oplus \Omega
_{M^{4}}^{1,1+}\oplus \Omega _{M^{4}}^{0,2},~~~~~\Omega _{M^{4}}^{2-}=\Omega
_{M^{4}}^{1,1-},  \label{decomposition_02}
\end{equation}%
where the superscripts $+$ and $-$ mean self-dual and anti-self-dual,
respectively. The two-forms $J_{M^{4}}$,$~\Omega _{M^{4}}$, and ${\bar{\Omega%
}}_{M^{4}}$ are in the spaces $\Omega _{M^{4}}^{1,1+},\Omega _{M^{4}}^{2,0}$
and $\Omega _{M^{4}}^{0,2}$ respectively. The condition (\ref{condition_})
implies that
\begin{eqnarray}
f_{(i)}\wedge J_{M^{4}} &=&0,~~~{\bar{f}}_{(i)}\wedge J_{M^{4}}=0,  \notag \\
f_{(i)}\wedge \Omega _{M^{4}} &=&0,~~~{\bar{f}}_{(i)}\wedge \Omega
_{M^{4}}=0.  \label{primitivity_01}
\end{eqnarray}%
This means that $f_{(i)}$,$~{\bar{f}}_{(i)}$ are perpendicular to $%
J_{M^{4}},\Omega _{M^{4}},{\bar{\Omega}}_{M^{4}}$, and are thus in the space
$\Omega _{M^{4}}^{1,1-}$. The first three equations imply the fourth
equation in (\ref{primitivity_01}), by the decomposition in (\ref%
{decomposition_02}). Since $f_{(i)},{\bar{f}}_{(i)}\in \Omega
_{M^{4}}^{1,1-} $,$~$the connections $a_{(i)},{\bar{a}}_{(i)}~$have
anti-self-dual curvatures, that is,
\begin{eqnarray}
f_{(i)} &=&-\ast _{4}f_{(i)}\,,  \notag \\
{\bar{f}}_{(i)} &=&-\ast _{4}{\bar{f}}_{(i)}.  \label{anti-self-duality}
\end{eqnarray}%
Connections with anti-self-dual curvatures on four-manifolds have been
discussed in, for example \cite{Donaldson,Atiyah}. The $f_{(i)},{\bar{f}}%
_{(i)}$ are of $(1,1)$ type here. Moreover, they are orthogonal to the
self-dual two forms. They are primitive $(1,1)$ forms. We hence refer to
Eqs. (\ref{primitivity_01}) and (\ref{condition_}) as primitivity condition.

The metric ansatz (\ref{M8_metric_01}) of $M^{8}$ fits into special cases of
eight-manifolds considered in \cite%
{Prins:2013koa,Rosa:2013lwa,Lau:2014fia,Minasian:2016txd}. Let us consider
the condition of $SU(4)$ structure for $M^{8}$. The $SU(4)$ structure
relation is given by
\begin{equation}
\frac{1}{2^{4}}\Omega \wedge {\bar{\Omega}=}\frac{1}{4!}J^{4},~~~~J\wedge
\Omega =0.  \label{SU(4) structure relation}
\end{equation}%
From the above ansatz (\ref{M8_metric_01}), (\ref{J_01}) and (\ref{Omega_01}%
),
\begin{eqnarray}
\frac{1}{2^{4}}\Omega \wedge {\bar{\Omega}} &=&\frac{1}{4}J_{(1)}\wedge
J_{(2)}\wedge \pi ^{\ast }(\Omega _{M^{4}}\wedge {\bar{\Omega}}%
_{M^{4}})e^{8v+4C}, \\
\frac{1}{4!}J^{4} &=&\frac{1}{2}J_{(1)}\wedge J_{(2)}\wedge \pi ^{\ast
}J_{M^{4}}^{2}e^{8v+4C}.
\end{eqnarray}%
The $SU(4)$ condition
\begin{equation}
\frac{1}{2^{4}}\Omega \wedge {\bar{\Omega}=}\frac{1}{4!}J^{4}
\label{SU(4)_01}
\end{equation}%
requires that
\begin{equation}
\frac{1}{2^{2}}\Omega _{M^{4}}\wedge {\bar{\Omega}}_{M^{4}}=\frac{1}{2!}%
J_{M^{4}}^{2},  \label{base_01}
\end{equation}%
and the $SU(4)$ condition
\begin{equation}
J\wedge \Omega =e^{6v+4C}\theta _{(1)}\wedge \theta _{(2)}\wedge \pi ^{\ast
}(J_{M^{4}}\wedge \Omega _{M^{4}})=0  \label{SU(4)_02}
\end{equation}%
requires that%
\begin{equation}
J_{M^{4}}\wedge \Omega _{M^{4}}=0.  \label{base_02}
\end{equation}%
The Eqs. (\ref{base_01}) and (\ref{base_02}) mean that the base $M^{4}$ has $%
SU(2)$ structure. Since $M^{4}$ is also K\"{a}hler, this means that $M^{4}~$%
is a Calabi-Yau two-fold. The general Calabi-Yau two-folds include both
compact Calabi-Yau two-folds such as K3 surfaces and non-compact Calabi-Yau
two-folds. In order that\ the model of $M^{8}$ has $SU(4)$ structure, the
base complex manifold $M^{4}$ is a Calabi-Yau two-fold. Moreover, we have
also showed that when the base $M^{4}$ is a Calabi-Yau two-fold, with the
additional assumption of the primitivity condition (\ref{primitivity_01}),
the $M^{8}$ is a conformally balanced $SU(4)$-structure Hermitian manifold.

\vspace{1pt}


\section{$G_{2}$ manifolds from the eight-manifolds}

\label{sec: G2}

In the previous section we have constructed a geometric model of
eight-dimensional manifolds by considering the manifold $M^{8}$ as a $T^{4}$
fibration over of a $M^{4}$ base. In this section we describe the $M^{8}$ in
another way. The $M^{8}$ can be viewed as a circle fibration over a
seven-dimensional manifold ${M}^{7}$,
\begin{equation}
S^{1}\overset{}{\rightarrow }M^{8}\overset{}{\rightarrow }M^{7}.
\end{equation}%
\vspace{1pt}We define the map
\begin{equation}
\tau :M^{8}\overset{}{\rightarrow }M^{7}.
\end{equation}%
According to the metric ansatz (\ref{M8_metric_01}), the $M^{7}$ is hence
the $T^{3}${\ fibration over }$M^{4},$\vspace{1pt}%
\begin{equation}
T^{3}\overset{}{\rightarrow }M^{7}\overset{}{\rightarrow }M^{4}
\end{equation}%
and we define the map
\begin{equation}
\psi :M^{7}\overset{}{\rightarrow }M^{4}.  \label{M7 projection}
\end{equation}%
The projection map $\pi $ in Sec. \ref{sec: T4} is hence
\begin{equation}
\pi =\psi \circ \tau .
\end{equation}%
As in Sec. \ref{sec: T4}, we consider the base $M^{4}$ to be a Calabi-Yau
two-fold, which has $SU(2)$ structure. We will see in this section that ${M}%
^{7}$ is a $G_{2}~$manifold with skew torsion.

A $G_{2}~$manifold with torsion contains a metric, a fundamental three-form$%
~\varphi _{3}$, and its dual four-form$~\varphi _{4}=\ast _{7}\varphi _{3}.$
If it has torsion, then$~d\varphi _{3}\neq 0,$ and the $d\varphi _{3}$
measures the torsion. For the classifications of $G_{2}~$manifolds with
torsion, see for example \cite%
{FG,Cabrera96,Swann96,Friedrich:2001nh,Friedrich:2001yp}.

The metric ansatz of the eight-manifold is
\begin{equation}
ds^{2}(M^{8})=e^{2v}(dx_{1}+\tau ^{\ast }\psi ^{\ast }\func{Re}%
a_{(1)})^{2}+ds^{2}({\tilde{M}}^{7})
\end{equation}%
where the $x_{1}$ parametrizes the coordinate of the $S^{1}$ and
\begin{equation}
ds^{2}({\tilde{M}}^{7})=e^{2v}ds^{2}({M}^{7}).
\end{equation}%
The seven-manifold we are looking at is%
\begin{eqnarray}
ds^{2}({M}^{7}) &=&(dy_{1}+\psi ^{\ast }\func{Im}a_{(1)})^{2}+(dx_{2}+\psi
^{\ast }\func{Re}a_{(2)})^{2}+(dy_{2}+\psi ^{\ast }\func{Im}%
a_{(2)})^{2}+e^{2C}ds^{2}(M^{4}),  \notag \\
&&
\end{eqnarray}%
where the $\{y_{1},x_{2},y_{2}\}$ are coordinates of the $T^{3}.$

We can define the fundamental three-form $\varphi _{3}$ of ${M}^{7}$,
\begin{eqnarray}
\varphi _{3} &=&(J_{(2)}+e^{2C}\psi ^{\ast }J_{M^{4}})\wedge (dy_{1}+\psi
^{\ast }\func{Im}a_{(1)})+e^{2C}\func{Im}(\theta _{(2)}\wedge \psi ^{\ast
}\Omega _{M^{4}})  \notag \\
&=&(dy_{1}+\psi ^{\ast }\func{Im}a_{(1)})\wedge (dx_{2}+\psi ^{\ast }\func{Re%
}a_{(2)})\wedge (dy_{2}+\psi ^{\ast }\func{Im}a_{(2)})  \notag \\
&&+e^{2C}[\psi ^{\ast }J_{M^{4}}\wedge (dy_{1}+\psi ^{\ast }\func{Im}%
a_{(1)})+\func{Im}(\theta _{(2)}\wedge \psi ^{\ast }\Omega _{M^{4}})].
\label{phi_3}
\end{eqnarray}%
The dual four-form $\varphi _{4}$ is
\begin{eqnarray}
\varphi _{4} &=&\ast _{7}\varphi _{3}  \notag \\
&=&\frac{1}{2}(J_{(2)}+e^{2C}\psi ^{\ast }J_{M^{4}})^{2}-e^{2C}\func{Re}%
(\theta _{(2)}\wedge \psi ^{\ast }\Omega _{M^{4}})\wedge (dy_{1}+\psi ^{\ast
}\func{Im}a_{(1)}).  \label{phi_4}
\end{eqnarray}

We see that
\begin{eqnarray}
d\varphi _{3} &=&\psi ^{\ast }\func{Im}f_{(2)}\wedge (dy_{1}+\psi ^{\ast }%
\func{Im}a_{(1)})\wedge (dx_{2}+\psi ^{\ast }\func{Re}a_{(2)})  \notag \\
&&+\psi ^{\ast }\func{Re}f_{(2)}\wedge (dy_{2}+\psi ^{\ast }\func{Im}%
a_{(2)})\wedge (dy_{1}+\psi ^{\ast }\func{Im}a_{(1)})  \notag \\
&&+\psi ^{\ast }\func{Im}f_{(1)}\wedge (dx_{2}+\psi ^{\ast }\func{Re}%
a_{(2)})\wedge (dy_{2}+\psi ^{\ast }\func{Im}a_{(2)})  \notag \\
&&+2e^{2C}dC\wedge \lbrack \psi ^{\ast }J_{M^{4}}\wedge (dy_{1}+\psi ^{\ast }%
\func{Im}a_{(1)})+\func{Im}(\theta _{(2)}\wedge \psi ^{\ast }\Omega
_{M^{4}})].  \label{d phi_3}
\end{eqnarray}%
Due to the non-zero $\psi ^{\ast }\func{Im}f_{(2)},\psi ^{\ast }\func{Re}%
f_{(2)},\psi ^{\ast }\func{Im}f_{(1)}~$in Eq. (\ref{d phi_3}), $\varphi _{3}$
is not closed, even if after a rescaling. In Eq. (\ref{d phi_3}), the
primitivity condition (\ref{primitivity_01}) has been used. We also see that
\begin{equation}
\varphi _{3}\wedge d\varphi _{3}=0,
\end{equation}%
in which we have used the primitivity condition and the $SU(2)$ structure
relation $J_{M^{4}}\wedge \Omega _{M^{4}}=0$. Meanwhile,
\begin{eqnarray}
d\varphi _{4} &=&2e^{2C}dC\wedge \lbrack J_{(2)}\wedge \psi ^{\ast
}J_{M^{4}}-\func{Re}(\theta _{(2)}\wedge \psi ^{\ast }\Omega _{M^{4}})\wedge
(dy_{1}+\psi ^{\ast }\func{Im}a_{(1)})]  \notag \\
&=&2e^{2C}dC\wedge \varphi _{4}  \notag \\
&=&{\hat{\theta}}\wedge \varphi _{4},
\end{eqnarray}%
where ${\hat{\theta}}$ is a Lee one-form
\begin{equation}
{\hat{\theta}}=\ 2e^{2C}dC.
\end{equation}%
Since
\begin{equation}
d\varphi _{4}={\hat{\theta}}\wedge \varphi _{4},~~d\varphi _{3}\neq
0,~~\varphi _{3}\wedge d\varphi _{3}=0,
\end{equation}%
this is a $G_{2}$ structure with skew torsion \cite%
{FG,Cabrera96,Swann96,Friedrich:2001nh,Friedrich:2001yp}. Since the Lee
one-form ${\hat{\theta}}~$is closed, that is $d{\hat{\theta}}=0$, this $%
G_{2} $ structure is locally conformal to a balanced $G_{2}$ structure, see
\cite{FG,Cabrera96,Swann96,Friedrich:2001nh,Friedrich:2001yp}. Hence, we see
that (${M}^{7},g_{{M}^{7}},\varphi _{3}$) constructed from the $T^{3}$
fibration over a Calabi-Yau two-fold gives a $G_{2}$ manifold with skew
torsion.

In the special case if ${\hat{\theta}}=\ 2e^{2C}dC$ vanish, then%
\begin{equation}
d\varphi _{4}=0,~~d\varphi _{3}\neq 0,~~\varphi _{3}\wedge d\varphi _{3}=0.
\end{equation}%
This is the condition of a balanced $G_{2}~$manifold \cite%
{FG,Cabrera96,Swann96,Friedrich:2001nh,Friedrich:2001yp,Fei:2015jxa}. Hence,
in this case, the seven-manifold ${M}^{7}~$is a balanced $G_{2}~$manifold.

In the above derivation, we have showed that $M^{8}$ is a circle fibration
over a $G_{2}$ manifold with skew torsion. The seven-manifold ${M}^{7},$ as
a $T^{3}$ bundle over $M^{4}$, can be described in another way. Let us
define the projection maps%
\begin{eqnarray}
\varrho &:&M^{7}\overset{}{\rightarrow }M^{6}, \\
\varsigma &:&M^{6}\overset{}{\rightarrow }M^{4},
\end{eqnarray}%
where ${M}^{6}$ is described by
\begin{equation}
ds^{2}({M}^{6})=(dx_{2}+\varsigma ^{\ast }\func{Re}a_{(2)})^{2}+(dy_{2}+%
\varsigma ^{\ast }\func{Im}a_{(2)})^{2}+e^{2C}ds^{2}(M^{4}),
\end{equation}%
\begin{equation}
J_{M^{6}}=J_{(2)}+e^{2C}\varsigma ^{\ast }J_{M^{4}},~~~~\Omega
_{M^{6}}=e^{2C}\theta _{(2)}\wedge \varsigma ^{\ast }\Omega _{M^{4}}.
\end{equation}%
The ${M}^{6}$ is a conformally balanced Hermitian manifold. Hence, the
projection map (\ref{M7 projection}) can be written as
\begin{equation}
\psi =\varsigma \circ \varrho .
\end{equation}%
Hence ${M}^{7}$ is also a circle fibration over $M^{6}.$ This circle is
parametrized by $y_{1}$. Considering it as a circle fibration of ${M}^{6},$
according to \cite{Chiossi}, we may also see that the Eqs. (\ref{phi_3}) and
(\ref{phi_4}) can also be written as
\begin{eqnarray}
\varphi _{3} &=&\varrho ^{\ast }J_{M^{6}}\wedge (dy_{1}+\varrho ^{\ast
}\varsigma ^{\ast }\func{Im}a_{(1)})+\varrho ^{\ast }\func{Im}(\Omega
_{M^{6}}), \\
\varphi _{4} &=&\frac{1}{2}\varrho ^{\ast }J_{M^{6}}\wedge \varrho ^{\ast
}J_{M^{6}}-\varrho ^{\ast }\func{Re}(\Omega _{M^{6}})\wedge (dy_{1}+\varrho
^{\ast }\varsigma ^{\ast }\func{Im}a_{(1)}).
\end{eqnarray}%
By an analysis similar to the one in Sec. \ref{sec: T4}, the base $M^{4}$
satisfies the $SU(2)$ structure relation (\ref{base_01}) and (\ref{base_02}%
). Hence the base $M^{4}$ has $SU(2)$ structure and we have showed in the
above that the $T^{3}$ bundle over $M^{4}$ has $G_{2}~$structure with skew
torsion.

\vspace{1pt}


\section{$SU(4)$ structures and fluxes}

\label{sec: type II string theory}

The previous sections have described the construction of the
eight-dimensional manifolds and their geometric properties. Let us now
discuss how these eight-manifolds can be used in string theory. Let us
consider to embed the metric ansatz (\ref{M8_metric_01}) of Sec. \ref{sec:
T4} in type II string theory. The ten-dimensional spacetime is a warped
product of a two-dimensional Minkowski spacetime and an eight-dimensional
manifold $M^{8}$, with the line element
\begin{equation}
ds^{2}=e^{2A}ds^{2}(R^{1,1})+ds^{2}(M^{8}),  \label{product_02}
\end{equation}%
where $e^{2A}$ is a warp factor in front of the metric of $R^{1,1}$. There
is also a dilaton field $\phi ~$in the ten-dimensional spacetime. The Poincar%
\'{e} invariance in two-dimensional Minkowski spacetime and the self-duality
constraint of the fluxes enables the decomposition \cite%
{Prins:2013koa,Rosa:2013lwa,Prins:2013wza} of the fluxes as
\begin{equation}
\mathcal{F}=\text{Vol}_{2}\wedge e^{2A}\ast _{8}\sigma F+F.  \label{flux_}
\end{equation}%
Here, Vol$_{2}$ is the volume form of $R^{1,1}.~$The $\mathcal{F}$ is a
polyform, which is the sum of the R-R fluxs of different ranks. The $F$ in
the ansatz (\ref{flux_}) is a polyform on the internal manifold. Let us
restrict our attention to type IIB, in which case $\mathcal{F}=\sum \mathcal{%
F}_{(k)}$, where $k=1,3,5,7,9$. The $\sigma $ is a sign factor, and $\sigma
\mathcal{F}_{(k)}=(-1)^{\frac{1}{2}k(k-1)}\mathcal{F}_{(k)}$ where $k$ is
the rank of the form. The self-duality constraint in type IIB theory is%
\begin{equation}
\mathcal{F}=\ast _{10}\sigma \mathcal{F,}  \label{self-duality constraint}
\end{equation}%
and is satisfied by the ansatz (\ref{flux_}).

The type IIB string theory in ten dimensions has two Killing spinors $%
\epsilon _{1},\epsilon _{2}~$of the same chirality. This case corresponds to
$\mathcal{N}=(2,0)$ supersymmetry in 1+1 dimensions. There are two positive
chirality supercharges, which can be denoted by a complex-valued Weyl spinor
$\zeta $ in 1+1 dimensions. For these solutions the most general
decomposition of the Killing spinors $\epsilon _{1},\epsilon _{2}\ $is given
by
\begin{eqnarray}
\epsilon _{1} &=&\zeta \otimes \eta _{1}+~\text{c.c.} \\
\epsilon _{2} &=&\zeta \otimes \eta _{2}+~\text{c.c.}\;
\end{eqnarray}%
where $\eta _{1},\eta _{2}$ are internal Killing spinors which are Weyl
spinors in 8 dimensions and they have the same chirality. The $M_{8}$ is
equipped with an $SU(4)$ structure, which is equivalent to the existence of
a pure spinor $\eta $. In this case, the pure spinor is $\eta \propto \eta
_{1}=e^{-i\vartheta }\eta _{2}$, up to a normalization factor. The pure
spinor $\eta ~$satisfies $\eta ^{t}\eta =0.$\vspace{1pt}~One can construct
an $SU(4)$ structure by taking the spinor bilinears \cite%
{Prins:2013koa,Rosa:2013lwa} of the internal Killing spinor%
\begin{eqnarray}
J_{mn} &=&-i\eta ^{\dagger }\gamma _{mn}\eta , \\
\Omega _{mnpq} &=&\eta ^{t}\gamma _{mnpq}\eta .
\end{eqnarray}%
By using Fierz identities one can show that these forms obey the $SU(4)$
structure relation (\ref{SU(4) structure relation}). The $J$ and $\Omega $
are the Hermitian two-form and holomorphic four-form respectively. The
polyforms can be written as%
\begin{eqnarray}
\Psi _{1} &=&-e^{-i\vartheta }e^{-iJ}, \\
\Psi _{2} &=&-e^{i\vartheta }\Omega ,
\end{eqnarray}%
where $e^{-i\vartheta }$ is a phase factor. For more discussions on the
properties of pure spinors, see for example \cite%
{Prins:2013koa,Rosa:2013lwa,Grana:2006kf,Tomasiello:2007zq,Andriot:2009fp,Halmagyi:2007ft,Lust:2010by}%
.

It can be shown \cite{Prins:2013koa,Rosa:2013lwa} that the supersymmetry
equations can be elegantly written with the pure spinors as%
\begin{eqnarray}
d_{H}\left( e^{2A-\phi }\text{Re}\Psi _{1}\right) &=&e^{2A}\ast \sigma F,
\label{pure_spinor_01} \\
d_{H}\left( e^{2A-\phi }\Psi _{2}\right) &=&0,  \label{pure_spinor_03} \\
i(\bar{\partial}_{H}-\partial _{H})\left( e^{-\phi }\text{Im}\Psi
_{1}\right) &=&F,  \label{pure_spinor_02}
\end{eqnarray}%
where $d_{H}$ is the twisted exterior derivative, and $d_{H}\equiv d+H\wedge
,$ where $H$ is the NS-NS three form. We can decompose it as $d_{H}=\partial
_{H}+\bar{\partial}_{H},~$where $\partial _{H}\equiv \partial
+H^{(2,1)}\wedge $ is the ordinary twisted Dolbeault operator and $\bar{%
\partial}_{H}=\bar{\partial}+H^{(1,2)}\wedge $ is its complex conjugate, and
$H^{(2,1)}$, $H^{(1,2)}~$are the $(2,1)$ type and $(1,2)$ type in $H.$ More
details on these equations and their generalizations have been discussed in
\cite{Prins:2013koa,Rosa:2013lwa}.

In the absence of $H$, $d_{H}$ reduces to $d$, and $i(\bar{\partial}%
_{H}-\partial _{H})$ reduces to $i(\bar{\partial}-\partial )=d^{c}.$\vspace{%
1pt} Let us consider also the absence of $F_{1}~$and $F_{5}.~$From the
differential equations for pure spinors (\ref{pure_spinor_01}) and (\ref%
{pure_spinor_02}), we have respectively
\begin{equation}
\ast _{8}F_{3}=\frac{1}{2}e^{-2A}d(e^{2A-\phi }J^{2})\,,  \label{*F3}
\end{equation}%
and%
\begin{equation}
F_{3}=i(\partial -\bar{\partial})(e^{-\phi }J)\,=-d^{c}(e^{-\phi }J).
\label{F3_01}
\end{equation}

Let us now combine the above two equations (\ref{*F3}) and (\ref{F3_01}),
and then we have
\begin{equation}
d^{c}(e^{-\phi }J)=\frac{1}{2}e^{-2A}\ast _{8}d(e^{2A-\phi }J^{2}).
\label{differential_02}
\end{equation}%
The case with constant $e^{\phi }$ and $e^{2A}$ is $d^{c}J=\frac{1}{2}\ast
_{8}d(J^{2})$, which was obtained in \cite{Minasian:2016txd}. Since $%
d(e^{2A-\phi }J^{2})$ appears in $F_{7}$, and $d^{c}(e^{-\phi }J)$ appears
in $F_{3},$ the Hodge dual relation (\ref{differential_02}) is closely
connected to the self-duality constraint (\ref{self-duality constraint}) in
the type IIB string theory.

The IIB theory contains the field equations \cite%
{Schwarz:1983qr,Bergshoeff:1995sq} in the string frame
\begin{equation}
d\ast ({\tilde{F}}_{3})=g_{s}F_{5}\wedge H_{3},
\end{equation}%
where ${\tilde{F}}_{3}=F_{3}+C_{0}H_{3},$ $F_{5}=dC_{4},~$and $C_{0}$ is the
axion. In the case without the axion and $F_{5},$ this equation reduces to$\
dF_{7}=0$ in our convention. This equation is equivalent to the Bianchi
identity from the pure spinor equations.

The Eq. (\ref{pure_spinor_03}) gives
\begin{equation}
d\left( e^{2A-\phi }\Omega \right) =0.  \label{holomorphic_02}
\end{equation}%
We use the ansatz (\ref{M8_metric_01}) for the eight-manifolds $M^{8}$ with $%
SU(4)$ structure which appear in the warped product (\ref{product_02}), in
the case of type IIB string theory. Comparing Eq. (\ref{holomorphic_01})
with Eq. (\ref{holomorphic_02}), we see that%
\begin{equation}
e^{2C}=e^{\phi -4v-2A}.  \label{relation_01}
\end{equation}%
In general, we can consider the eight-manifolds $M^{8}$\ to be either
compact or non-compact. For instance, we can obtain non-compact $M^{8}$ by
taking the base $M^{4}$ in the ansatz (\ref{M8_metric_01}) to be
non-compact. The eight-dimensional non-compact models in the ten-dimensional
string theory may be considered as local models of compact solutions.

If demanding $d\Omega =0$, which means that $M^{8}$ has an integrable
complex structure, we have
\begin{equation}
e^{\phi }=e^{2A}.  \label{e^2A_}
\end{equation}%
However, this equation (\ref{e^2A_}) should correspond to a special case,
which should lead to special solutions.

The $F_{3}$ is%
\begin{eqnarray}
F_{3} &=&-d^{c}(e^{-\phi }J)=i(\partial -\bar{\partial})(e^{2v-\phi
}(J_{(1)}+J_{(2)}+e^{2C}\pi ^{\ast }J_{M^{4}})) \\
&=&e^{2v-\phi }(\func{Re}({\bar{F}}_{(1)}\wedge \theta _{(1)})+\func{Re}({%
\bar{F}}_{(2)}\wedge \theta _{(2)}))  \notag \\
&&+i(\partial -\bar{\partial})(e^{2v-\phi })\wedge
(J_{(1)}+J_{(2)})+i(\partial -\bar{\partial})(e^{2v-\phi +2C})\wedge \pi
^{\ast }J_{M^{4}}.  \label{F3}
\end{eqnarray}%
The $F_{3}$ contains a nonzero piece $ie^{-\phi }(\partial -\bar{\partial})J$%
, hence the non-K\"{a}hlerity of the eight-manifold $M^{8}$ and the
non-closure of $J$ is closely related to the $F_{3}$. Acting on $F_{3}$ by a
further $d$,
\begin{eqnarray}
dF_{3} &=&-dd^{c}(e^{-\phi }J)  \notag \\
&=&e^{2v-\phi }({\bar{F}}_{(1)}\wedge F_{(1)}+{\bar{F}}_{(2)}\wedge F_{(2)})
\notag \\
&&+d(e^{2v-\phi })\wedge \left( i(\partial -\bar{\partial}%
)(J_{(1)}+J_{(2)})\right)  \notag \\
&&-2i\partial \bar{\partial}(e^{2v-\phi })\wedge
(J_{(1)}+J_{(2)})-2i\partial \bar{\partial}(e^{2v-\phi +2C})\wedge \pi
^{\ast }J_{M^{4}}.
\end{eqnarray}

The $F_{7}$ is
\begin{equation}
F_{7}=\frac{1}{2}e^{2A}\text{Vol}_{R^{1,1}}\wedge e^{-2A}d(e^{2A-\phi
}J^{2})=\frac{1}{2}\text{Vol}_{R^{1,1}}\wedge d(e^{2A-\phi }J^{2}).
\label{F7}
\end{equation}%
According to Eq. (\ref{dJ^2}) in Sec. \ref{sec: T4},
\begin{eqnarray}
d(e^{2A-\phi }J^{2}) &=&-e^{4v+2A-\phi }(i\func{Im}(F_{(1)}\wedge {\bar{%
\theta}}_{(1)})\wedge \theta _{(2)}\wedge {\bar{\theta}}_{(2)}+i\func{Im}%
(F_{(2)}\wedge {\bar{\theta}}_{(2)})\wedge \theta _{(1)}\wedge {\bar{\theta}}%
_{(1)}))  \notag \\
&&+e^{2A-\phi }(d(4v+2A-\phi )\wedge J^{2}+4e^{4v+2C}dC\wedge \pi ^{\ast
}J_{M^{4}}\wedge (J_{(1)}+J_{(2)})).  \notag \\
&&
\end{eqnarray}%
Using the relation (\ref{relation_01}), this can be simplified to
\begin{eqnarray}
d(e^{2A-\phi }J^{2}) &=&-e^{4v+2A-\phi }(i\func{Im}(F_{(1)}\wedge {\bar{%
\theta}}_{(1)})\wedge \theta _{(2)}\wedge {\bar{\theta}}_{(2)}+i\func{Im}%
(F_{(2)}\wedge {\bar{\theta}}_{(2)})\wedge \theta _{(1)}\wedge {\bar{\theta}}%
_{(1)}))  \notag \\
&&-4e^{4v+2A-\phi }dC\wedge J_{(1)}\wedge J_{(2)}.  \label{J^2_02}
\end{eqnarray}

We use an identity%
\begin{equation}
\ast _{8}((e^{2v+2C}F_{(i)})\wedge (e^{v}\func{Re}\theta _{(1)})\wedge
(e^{2v}\frac{i}{2}\theta _{(2)}\wedge {\bar{\theta}}%
_{(2)}))=-(e^{2v+2C}F_{(i)})\wedge (e^{v}\func{Im}\theta _{(1)}),
\label{identity_01}
\end{equation}%
in which the anti-self-duality (\ref{anti-self-duality}) has been used.
After using Eq. (\ref{identity_01}), we see that the pieces in (\ref{F3})
and (\ref{J^2_02}) involving $F_{(i)}$ are satisfied for the Eq. (\ref%
{differential_02}). Let us now look at the pieces in (\ref{F3}) and (\ref%
{J^2_02}) which do not involve $F_{(i)}$. By comparing these pieces in Eq. (%
\ref{differential_02}), we see that
\begin{equation}
(\partial -\bar{\partial})(e^{2v-\phi })=0,
\end{equation}%
\begin{equation}
i(\partial -\bar{\partial})(e^{2v-\phi +2C})\wedge \pi ^{\ast
}J_{M^{4}}=\ast _{8}(2e^{4v-\phi }dC\wedge J_{(1)}\wedge J_{(2)}).
\end{equation}%
Hence, from the first equation above we have that
\begin{equation}
e^{2v}=e^{\phi }.  \label{relation_02}
\end{equation}%
The second equation becomes
\begin{equation}
i(\partial -\bar{\partial})(e^{2C})\wedge (e^{2v+2C}\pi ^{\ast
}J_{M^{4}})=\ast _{8}(e^{4v}d(e^{2C})\wedge J_{(1)}\wedge J_{(2)}).
\end{equation}%
Using the metric (\ref{M8_metric_01}) and a similar identity as (\ref%
{identity_01}),
\begin{equation}
d^{c}(e^{2C})\wedge J_{M^{4}}=\ast _{4}d(e^{2C}).
\end{equation}%
Acting on both sides by $d$,
\begin{equation}
2i\partial \bar{\partial}(e^{2C})\wedge J_{M^{4}}=\frac{1}{2}\Delta
(e^{2C})J_{M^{4}}\wedge J_{M^{4}},
\end{equation}%
where $\Delta $ is the Laplacian. Hence we have that
\begin{eqnarray}
dF_{3} &=&e^{2v-\phi }({\bar{F}}_{(1)}\wedge F_{(1)}+{\bar{F}}_{(2)}\wedge
F_{(2)})-2i\partial \bar{\partial}(e^{2v-\phi +2C})\wedge \pi ^{\ast
}J_{M^{4}}  \notag \\
&=&{\bar{F}}_{(1)}\wedge F_{(1)}+{\bar{F}}_{(2)}\wedge F_{(2)}-\frac{1}{2}%
\Delta (e^{2C})\pi ^{\ast }J_{M^{4}}^{2},
\end{eqnarray}%
where we have used the relation (\ref{relation_02}). Using the relations (%
\ref{relation_01}), (\ref{relation_02}) and the condition (\ref{e^2A_}), we
have that
\begin{equation}
e^{2C}=e^{-2\phi }.
\end{equation}

Because of the anti-self-duality of the curvatures $f_{(1)}$ and $f_{(2)},~$
\begin{equation}
{\bar{f}}_{(i)}\wedge f_{(i)}=-{\bar{f}}_{(i)}\wedge \ast
_{4}f_{(i)}=-|f_{(i)}|^{2}\text{Vol}_{4}=-\frac{1}{2}%
|f_{(i)}|^{2}J_{M^{4}}^{2},  \label{f_i_05}
\end{equation}%
for$~i=1,2$, where Vol$_{4}$ is the volume form of $M^{4}$ and we used the
anti-self-duality $f_{(i)}=-\ast _{4}f_{(i)}^{4}$.~Hence,
\begin{equation}
{\bar{F}}_{(i)}\wedge F_{(i)}=-\frac{1}{2}|f_{(i)}|^{2}\pi ^{\ast
}J_{M^{4}}^{2}.
\end{equation}

The Bianchi identity for the $F_{3}$ flux, in the presence of D5 and O5
sources, is
\begin{equation}
dF_{3}=\rho ^{(4)}(D5)-\rho ^{(4)}(O5),  \label{source_01}
\end{equation}%
where $\rho ^{(4)}(D5)$ and $\rho ^{(4)}(O5)$ are the four-form Poincar\'{e}
duals to the four-cycles that D5 and O5 wraps inside of $M^{8}$. The D5 is
positively charged and the O5 is negatively charged. They are the source
terms for Eq. (\ref{source_01}).\vspace{1pt} In the case that $e^{2v-\phi }$
is constant,
\begin{equation}
dF_{3}=-\frac{1}{2}|f_{(1)}|^{2}\pi ^{\ast }J_{M^{4}}^{2}-\frac{1}{2}%
|f_{(2)}|^{2}\pi ^{\ast }J_{M^{4}}^{2}-\frac{1}{2}\Delta (e^{2C})\pi ^{\ast
}J_{M^{4}}^{2}.  \label{source_02}
\end{equation}%
Hence, to balance the right hands of (\ref{source_01}) and (\ref{source_02}%
), we have
\begin{equation}
{\bar{F}}_{(1)}\wedge F_{(1)}+{\bar{F}}_{(2)}\wedge F_{(2)}-\frac{1}{2}%
\Delta (e^{-2\phi })\pi ^{\ast }J_{M^{4}}^{2}=\rho ^{(4)}(D5)-\rho
^{(4)}(O5).  \label{tadpole_cancellation}
\end{equation}%
This is a tadpole cancellation condition. On the right hand of Eq. (\ref%
{tadpole_cancellation}), for general configurations, we have considered the
inclusion of the negatively charged O5. The existence of negatively charged
O5 has been anticipated in \cite%
{Grana:2006kf,Tomasiello:2007zq,Andriot:2009fp}.

This is a configuration with $F_{3}$ and $F_{7}$ fluxes and dilaton, in the
warped product (\ref{product_02}) of two-dimensional Minkowski spacetime and
$SU(4)$-structure eight-manifold $M^{8}$ with the metric (\ref{M8_metric_01}%
). The geometric model constructed in Sec. \ref{sec: T4} is hence realized
in the type IIB string theory.

\vspace{1pt}


\section{Generalized calibrated cycles}

\label{sec: generalized calibrated cycles}

In the previous sections, we have discussed the geometric model of the
eight-manifolds and their realizations in type II string theory. Now we
discuss more about specific geometric structures on these manifolds. A
natural set of geometric structures are calibrated cycles, and in particular
the generalized calibrated cycles in the presence of fluxes.

In the type II string theory, branes can wrap calibrated cycles. These
cycles are calibrated by calibration forms. A usual calibration form is a
closed form, and when restricted to the calibrated cycle, is the volume form
\cite{Harvey:1982xk}. There exist generalized calibration forms in the
presence of background fluxes. These generalized calibration forms are not
closed, due to the fluxes. However, the generalized calibration forms
twisted by background form-potentials are closed. Meanwhile, the brane
actions contain two types of terms. One type is the pull-back of volume, and
another type in the brane action is the pull-back of background
form-potentials, for example the R-R potentials. The generalized
calibrations after subtracting the pull-back of background form-potentials,
are hence closed. For more discussions on generalized calibrations, see for
example \cite{Gutowski:1999tu,Gauntlett:2003di,Lust:2010by} and references
therein.

In the geometric model constructed in Sec. 2, we may denote the torus
parametrized by $x_{1},y_{1}$ as $T_{(1)}^{2}$, and the torus parametrized
by $x_{2},y_{2}$ as $T_{(2)}^{2}$. Consider a K\"{a}hler two-cycle $\Sigma
_{(1)}^{2}$ inside the base $M^{4},$ calibrated by $J_{M^{4}}$, so $%
J_{M^{4}}|_{\Sigma _{(1)}^{2}}=\mathrm{Vol}_{\Sigma _{(1)}^{2}},$ which is
the volume form of the two-cycle. We can make a four-cycle $\Sigma
_{(1)}^{4} $, which is the restriction of the $T_{(1)}^{2}$ fibration to the
submanifold $\Sigma _{(1)}^{2}\subset M^{4}.~$ Similarly, we can consider a K%
\"{a}hler two-cycle $\Sigma _{(2)}^{2}$ $\subset $ $M^{4},$ calibrated by $%
J_{M^{4}},$ so that $J_{M^{4}}|_{\Sigma _{(2)}^{2}}=\mathrm{Vol}_{\Sigma
_{(2)}^{2}}.~$In the similar way, we can make a four-cycle $\Sigma
_{(2)}^{4} $, which is the restriction of the $T_{(2)}^{2}$ fibration
instead, to the submanifold $\Sigma _{(2)}^{2}$ inside $M^{4}.~$

Let us consider that the fivebrane is parallel to $R^{1,1}$ and wraps a
four-cycle $\Sigma ^{4}$ inside $M^{8}$. The worldvolume of the fivebrane is
hence $R^{1,1}\times \Sigma ^{4}$. In the case at hand, the brane action is $%
S=-\mu _{5}\int d^{6}\sigma e^{-\phi }\sqrt{-\det g_{\parallel }}\sqrt{\det
g_{\perp }}+\mu _{5}\int C_{6},$ in which $g_{\parallel }~$is the induced
worldvolume metric on the $R^{1,1}$ directions where$~\sqrt{-\det
g_{\parallel }}=e^{2A}$, and\ $g_{\perp }~$is the induced worldvolume metric
on $\Sigma ^{4}$. The $d^{6}\sigma $ is the volume element and the $\mu _{5}$
is the charge of the brane. The brane configuration on the generalized
calibrated cycle minimizes the total energy. This total energy is the sum of
the energy coming from the tension on the worldvolume and that coming from
the coupling of the brane to the background form-potential. The background
form-potential here is $C_{6}$. We can write it as
\begin{equation}
C_{6}=e^{2A-\phi }\mathrm{Vol}_{R^{1,1}}\wedge \Pi _{4}.
\end{equation}%
The energy density of the fivebrane on $\Sigma ^{4}$ is $E,$ and
\begin{equation}
\int d^{4}\sigma ~E=\int d^{4}\sigma ~e^{2A-\phi }\sqrt{\det g_{\perp }}%
-\int e^{2A-\phi }\Pi _{4},
\end{equation}%
where $d^{4}\sigma $ is the volume element of $\Sigma ^{4}$. General
discussions on calibrations on $SU(4)$-structure manifolds have been
considered in \cite{Prins:2013koa}. The generalized calibration form $%
e^{2A-\phi }\Xi _{4}$, for any cycle $\Sigma ^{\prime 4}$, satisfies the
inequality
\begin{equation}
e^{2A-\phi }\Xi _{4}|_{\Sigma ^{\prime 4}}\leq e^{2A-\phi }d^{4}\sigma \sqrt{%
\det g_{\perp }}|_{\Sigma ^{\prime 4}}  \label{generalized calibration_02}
\end{equation}%
where $d^{4}\sigma \sqrt{\det g_{\perp }}|_{\Sigma ^{\prime 4}}$ is the
volume form of $\Sigma ^{\prime 4},$ and the equality is satisfied for
calibrated cycles. From the Eq. (\ref{F7}), the $F_{7}$ is
\begin{equation}
F_{7}=\text{Vol}_{R^{1,1}}\wedge d(e^{2A-\phi }\frac{1}{2}J\wedge J).
\end{equation}%
We also see from the Eq. (\ref{holomorphic_02}) that
\begin{equation}
d(e^{2A-\phi }\func{Re}(e^{i\beta }\Omega ))=0,
\end{equation}%
where$~e^{i\beta }$ is a phase factor. Since $F_{7}-dC_{6}=0$, we have that
\begin{equation}
d[\text{\textrm{Vol}}_{R^{1,1}}\wedge e^{2A-\phi }\Xi _{4}-C_{6}]=0,
\label{form_01}
\end{equation}%
where%
\begin{equation}
\Xi _{4}=\frac{1}{2}J\wedge J+\func{Re}(e^{i\beta }\Omega ).
\end{equation}%
This agrees with the observations in \cite{Prins:2013koa}. This means that $%
e^{2A-\phi }\mathrm{Vol}_{R^{1,1}}\wedge \Xi _{4}$, after subtracting $C_{6}$%
, is closed. Hence, we can identify the generalized calibration form in the
ten dimensions as
\begin{eqnarray}
\Xi _{6} &=&\mathrm{Vol}_{R^{1,1}}\wedge e^{2A-\phi }\Xi _{4}  \notag \\
&=&e^{2A-\phi }\mathrm{Vol}_{R^{1,1}}\wedge (\frac{1}{2}J\wedge J+\func{Re}%
(e^{i\beta }\Omega )).
\end{eqnarray}%
The generalized calibration form twisted by the background form-potential is
\begin{equation}
\Xi _{6}^{\prime }=\Xi _{6}-C_{6}.
\end{equation}%
According to Eq. (\ref{form_01}), the $\Xi _{6}$ after subtracting the
background form-potential,~is closed, that is, $d\Xi _{6}^{\prime }=0$.

Inside $M^{8},$ the above generalized calibration form corresponds to$%
~e^{2A-\phi }\Xi _{4}$. This form after subtracting the background
form-potential $e^{2A-\phi }\Pi _{4}$ is
\begin{equation}
\Xi _{4}^{\prime }=e^{2A-\phi }\Xi _{4}-e^{2A-\phi }\Pi _{4},
\end{equation}%
and $d\Xi _{4}^{\prime }=0.~$The $M^{8}$ here is not a usual K\"{a}hler
Calabi-Yau four-fold. In the case of the K\"{a}hler Calabi-Yau four-fold,
the calibrations have been considered in \cite{Becker:1996ay}. According to
Eq. (\ref{generalized calibration_02}), the restriction of the $\Xi _{4}$ to
a calibrated cycle is the volume form of the calibrated cycle. The
restriction of the $\Xi _{4}$ to the four-cycle $\Sigma _{T}^{4}=T^{4}$ is
\begin{equation}
\Xi _{4}|_{\Sigma _{T}^{4}}=-\frac{1}{4}e^{4v}\theta _{(1)}\wedge {\bar{%
\theta}}_{(1)}\wedge \theta _{(2)}\wedge {\bar{\theta}}_{(2)}|_{\Sigma
_{T}^{4}},
\end{equation}%
which is the volume form $e^{4v}dx_{1}\wedge dy_{1}\wedge dx_{2}\wedge
dy_{2}~$of the cycle $\Sigma _{T}^{4}~$inside $M^{8}$.~This means that $%
\Sigma _{T}^{4}$ is a generalized calibrated cycle. As constructed above,
the four-cycle $\Sigma _{(1)}^{4}$ is the restriction of the $T_{(1)}^{2}$
fibration over the K\"{a}hler two-cycle $\Sigma _{(1)}^{2}$ in $M^{4}.~$The
restriction of the $\Xi _{4}$ to the four-cycle $\Sigma _{(1)}^{4}$ is%
\begin{equation}
\Xi _{4}|_{\Sigma _{(1)}^{4}}=e^{4v+2C}(\frac{i}{2}\theta _{(1)}\wedge {\bar{%
\theta}}_{(1)})\wedge \,\pi ^{\ast }J_{M^{4}}|_{\Sigma _{(1)}^{4}},
\end{equation}%
which is the volume form of the cycle $\Sigma _{(1)}^{4}~$inside $M^{8}.$
Hence $\Sigma _{(1)}^{4}$ is a generalized calibrated cycle. Similarly, the
four-cycle $\Sigma _{(2)}^{4}$ is the restriction of the $T_{(2)}^{2}$
fibration over the K\"{a}hler two-cycle $\Sigma _{(2)}^{2}$ in $M^{4}.~$The
restriction of the $\Xi _{4}~$to the four-cycle $\Sigma _{(2)}^{4}$ is%
\begin{equation}
\Xi _{4}|_{\Sigma _{(2)}^{4}}=e^{4v+2C}(\frac{i}{2}\theta _{(2)}\wedge {\bar{%
\theta}}_{(2)})\wedge \,\pi ^{\ast }J_{M^{4}}|_{\Sigma _{(2)}^{4}},
\end{equation}%
which is the volume form of the cycle $\Sigma _{(2)}^{4}~$inside $M^{8}$.
This shows that $\Sigma _{(2)}^{4}$ is a generalized calibrated cycle.
Hence, we have showed that the above generalized calibration form, after
subtraction of the background form-potential, is closed, and that when
restricted to the generalized calibrated cycle, is the volume form.

\vspace{1pt}

\vspace{1pt}

\vspace{1pt}


\section{Discussion}

\label{sec: discussion}

\vspace{1pt}

The geometric model of $T^{4}$ fibrations over Calabi-Yau two-folds
constructed in this paper provides examples of eight-dimensional balanced
manifolds and non-K\"{a}hler Hermitian manifolds. A seven-dimensional $G_{2}$
manifold with torsion also occurs in the construction of the present paper.
The eight-manifold of this type can also be viewed as a circle bundle over a
$G_{2}$ manifold with skew torsion. The eight-manifolds constructed here are
used in ten-dimensional models in type IIB string theory. These models have
rich geometric structures, such as fluxes and generalized calibrated cycles.

\vspace{1pt}

The IIB configurations here have similarities with configurations in
heterotic string theory. The $F_{3}~$flux in the type IIB case plays similar
role as the $H_{3}~$flux in the heterotic theory. The anomaly cancellation
condition in the heterotic case can be viewed as a counterpart to the
tadpole condition for fivebranes in the type IIB case here. The Hermitian
Yang-Mills equations \cite{Donaldson,UY} in the heterotic case, are
counterparts to the generalized calibrations in the type IIB case at hand.

\vspace{1pt}

The $T^{4}$ fibrations over Calabi-Yau two-folds considered here can also be
used as background manifolds for heterotic string theory. In the case for
heterotic theory, we need also to consider vector-bundles on these
eight-manifolds. One can construct stable vector bundles over the
eight-manifolds by pulling back stable bundles over the $\mathrm{{CY}_{2}}$
base space. Various methods of constructing vector-bundles in heterotic
theory for six-manifolds may be used for eight-manifolds.

\vspace{1pt}

The non-K\"{a}hler geometries considered here would be useful for mirror
symmetry for eight-dimensional non-K\"{a}hler manifolds \cite{Lau:2014fia}.
The $T^{4}$ fibration is analogous to the $T^{3}$ in SYZ proposal \cite%
{Strominger:1996it}, but for non-K\"{a}hler backgrounds. It may be
interesting to perform T-duality transformations along $T^{4}$. The examples
in this paper may serve as useful examples for performing T-dualities \cite%
{Bouwknegt:2003zg,Israel:2013hna} along higher dimensional tori.

\vspace{1pt}

It would be interesting to add $H_{3}$ flux in the IIB case here and obtain
more general configurations. In the presence of the $H_{3}$ flux, the
Dolbeault operators become twisted Dolbeault operators, which are twisted by
$H_{3}$ \cite{Prins:2013koa,Rosa:2013lwa,Tomasiello:2007zq}. We leave these
interesting and more general cases for future investigations.

\section*{Acknowledgments}

We would like to thank T. Fei, I. Melnikov, R. Minasian, D. Prins, D. Rosa,
V. Tosatti, C.-L. Wang, B. Wu, and S.-T. Yau for discussions and
communications. The work was supported in part by NSF grant DMS-1159412, NSF
grant PHY-0937443 and NSF grant DMS-0804454, and in part by YMSC, Tsinghua
University.

\vspace{1pt}

\vspace{1pt}


\end{document}